# Enhanced predictions of the Madden-Julian oscillation using the FuXi-S2S machine learning model: Insights into physical mechanisms

Can Cao[a,b], Xiaohui Zhong[c,d], Lei Chen[c,d], Zhiwei Wu[a,b],*, Hao Li[c,d],*

[a] *Department of Atmospheric and Oceanic Sciences and Institute of Atmospheric Sciences / Key Laboratory of Polar Atmosphere-ocean-ice System for Weather and Climate, Ministry of Education / Shanghai Scientific Frontier Base of Ocean-Atmosphere Interaction, Fudan University, Shanghai, China*
[b] *IRDR ICoE on Risk Interconnectivity and Governance on Weather/Climate Extremes Impact and Public Health, Fudan University, Shanghai, China*
[c] *Artificial Intelligence Innovation and Incubation Institute, Fudan University, Shanghai, China.*
[d] *Shanghai Academy of Artificial Intelligence for Science, Shanghai, China*
\* *Corresponding author.*
*E-mail address:* zhiweiwu@fudan.edu.cn *(Zhiwei Wu);* lihao_lh@fudan.edu.cn*(Hao Li).*

## ABSTRACT

The Madden-Julian Oscillation (MJO) is the dominant mode of tropical atmospheric variability on intraseasonal timescales, and reliable MJO predictions are essential for protecting lives and mitigating impacts on societal assets. However, numerical models still fall short of achieving the theoretical predictability limit for the MJO due to inherent constraints. In an effort to extend the skillful prediction window for the MJO, machine learning (ML) techniques have gained increasing attention. This study examines the MJO prediction performance of the FuXi subseasonal-to-seasonal (S2S) ML model during boreal winter, comparing it with the European Centre for Medium-Range Weather Forecasts S2S model. Results indicate that for the initial strong MJO phase 3, the FuXi-S2S model demonstrates reduced biases in intraseasonal outgoing longwave radiation anomalies averaged over the tropical western Pacific (WP) region during days 15-20, with the convective center located over this area. Analysis of multi-scale interactions related to moisture transport suggests that improvements could be attributed to the FuXi-S2S model's more accurate prediction of the area-averaged meridional gradient of low-frequency background moisture over the tropical WP. These findings not only explain the enhanced predictive capability of the FuXi-S2S model but also highlight the potential of ML approaches in advancing the MJO forecasting.

## 1. Introduction

The Madden-Julian oscillation (MJO) [1] is a prominent mode globally on intraseasonal timescales during boreal winter, characterized as a planetary-scale, coupled convection-circulation system that propagates eastward from the Indian Ocean (IO) to the western Pacific (WP) at an averaged phase speed of ~5 m/s. As it moves eastward, the MJO-related diabatic heating anomaly interacts with the mean atmospheric state to excite Rossby waves and influence the extratropical weather and climate [2-4]. For instance, extensive studies have demonstrated that MJO could effectively explain near-surface air temperature and precipitation anomalies over North America and East Asia [2,5-7], Tibetan Plateau snow cover [8], the North Atlantic oscillation [9,10], the Arctic oscillation [11], the Pacific-North America pattern [12], stratospheric sudden warming events [13], Antarctic sea ice [14], storm tracks [15,16] and atmospheric rivers [17-19]. Given its substantial impacts on global atmospheric intraseasonal variability, the MJO is widely recognized as the leading source of the global subseasonal prediction, bridging the gap between traditional weather forecasts (i.e., from 1 day to 2 weeks) and seasonal forecasts (i.e., from 2 months to 1 year) [20].

Recent advancements in theoretical frameworks [21], such as the skeleton theory [22], moisture-mode theory [23], gravity-wave theory [24] and trio-interaction theory [25], along with improvements in numerical models, have moderately advanced the MJO forecasting. Currently, state-of-the-art subseasonal-to-seasonal (S2S) models could predict the MJO up to 3-4 weeks in advance [26-28]. However, numerical models still fall short of achieving the theoretical predictability limit for the MJO, approximately 7 weeks, due to inherent constraints [29]. Notably, most models underestimate MJO signals over the WP region compared to observations, especially when the MJO convective center initiates over the IO, crosses the Maritime Continent (MC), and moves further into the WP, a phenomenon known as the MC prediction barrier [28,30]. Consequently, global subseasonal forecasting remains a grand challenge.

To extend the MJO's skillful prediction window, machine learning (ML) approaches have been employed, including direct forecasting of variables [31,32] and bias correction in post-processing [33]. Recent work has revealed that the FuXi S2S ML model (FuXi-S2S model for convenience), a machine learning model, extends the skillful MJO prediction from 30 to 36 days, surpassing the traditional European Centre for Medium-Range Weather Forecasts (ECMWF) S2S model [32]. However, this prior research provides only a preliminary examination of the MJO performance in the FuXi-S2S model. Therefore, in this study, we systematically compare ensemble mean forecasts of the MJO during boreal winter between the FuXi-S2S and ECMWF S2S models by analyzing model results spanning from 2017 to 2021. We analyze results from 2017 to 2021 because the FuXi-S2S model relies on 67-year data ranging from 1950 to 2016, and forecast results from 2017 to 2021 are used for evaluation. Additionally, we investigate whether the FuXi-S2S model could capture fundamental physical processes more accurately.

## 2. Materials and methods

### 2.1. Materials

*Reanalysis Datasets*

Reanalysis datasets used in this study contain daily mean moisture and horizontal winds at different pressure levels with a spatial resolution of 1° × 1° from the ERA5 and OLR data with a 2.5° × 2.5° horizontal resolution from the National Oceanic and Atmospheric Administration.

*Model outputs*

The FuXi-S2S model forecasts 76 variables, including 5 upper-air atmospheric variables at 13 pressure levels (50, 100, 150, 200, 250, 300, 400, 500, 600, 700, 850, 925, and 1000 hPa) and 11 surface variables. The model is trained using a 67-year dataset spanning from 1950 to 2016, and forecast results from 2017 to 2021 are used for evaluation. More details about the FuXi-S2S model can be found in Chen et al. [32]. For comparison, we select ECMWF S2S reforecasts generated from the model cycle C47r3. These reforecasts encompass initialization dates over 20 years, from January 3, 2002, to December 29, 2021, initialized twice weekly. It is worthwhile mentioning that the FuXi-S2S model has 51 members while the ECMWF S2S model has 11 members. Prediction capabilities are evaluated based on ensemble mean results. To ensure an equitable comparison, FuXi-S2S forecasts are selected for the same initialization dates as those utilized for ECMWF S2S reforecasts. In this study, we focus on the boreal winter, November (0) - April (1) (NDJMFA). Moisture, horizontal winds and OLR data from ECMWF S2S reforecasts and FuXi-S2S forecasts are used. All variables from reanalysis datasets and model outputs are interpolated into a 2.5° × 2.5° horizontal resolution.

*2.2. Methods*

With the purpose of monitoring the MJO activity, the Real-time Multivariate MJO (RMM) index is proposed by Wheeler and Hendon [34] (denoted as WH04), which has been widely used to evaluate the model forecasting capability of the MJO. Following the methods outlined in WH04, the RMM index for observation is calculated in three steps. First, datasets for OLR, zonal winds at 200 and 850 hPa spanning from 2002 to 2016 are used, and seasonal cycles (mean and the first three harmonics of the daily climatology for 2002-2016) are first eliminated. Then, the previous 120-day mean is subtracted to remove the interannual variability, decadal variability, and trend. Finally, multivariate empirical orthogonal functions (EOFs) are calculated for normalized daily anomalies of combined OLR, zonal winds at 200 and 850 hPa, averaged over 15°S to 15°N. The time series of the first two principal components (RMM1 and RMM2) obtained here are highly correlated with results downloaded from the Australian Bureau of Meteorology website (http://www.bom.gov.au/climate/mjo/). The amplitude of the RMM index is defined as the $\sqrt{RMM1^2+RMM2^2}$. And the RMM index during 2017-2021 in observation is achieved by projecting anomalies of OLR, 200- and 850-hPa zonal winds onto previous EOFs we get. For FuXi-S2S forecasts and ECMWF-S2S reforecasts, predicted RMM indexes are calculated in the following three steps. First, daily anomalies are calculated by removing the model climatology, defined for 2002-2016, which is a function of both the initial calendar date and lead time. Then, interannual, decadal and trend components are eliminated by subtracting the previous 120-day mean, which combines observations and subsequent model forecasts before the target date. Finally, predicted RMM indexes for both models are further obtained by projecting these anomalies, which are normalized by observational standard deviations, onto the first two observational EOF patterns.

To evaluate the MJO prediction skill, the anomaly correlation coefficient (ACC) and root-mean-square error (RMSE) are calculated [35,36]. These indices are defined as follows:

$$ACC(\tau) = \frac{\sum_{i=1}^{N}[a_{1i}(t)b_{1i}(t)+a_{2i}(t)b_{2i}(t)]}{\sqrt{\sum_{i=1}^{N}[a_{1i}^2(t)+a_{2i}^2(t)]}\sqrt{\sum_{i=1}^{N}[b_{1i}^2(t)+b_{2i}^2(t)]}} \qquad (1)$$

$$RMSE(\tau) = \sqrt{\frac{1}{N}\sum_{i=1}^{N}\{[a_{1i}(t) - b_{1i}(t)]^2 + [a_{2i}(t) - b_{2i}(t)]^2\}} \qquad (2)$$

where $a_{1i}(t)$ and $a_{2i}(t)$ are observed RMM1 and RMM2 at time $t$, and $b_{1i}(t, \tau)$ and $b_{2i}(t, \tau)$ are model predicted RMM1 and RMM2 with a $\tau$-day lead. $N$ is the total number of forecasts.

To provide a physical explanation of model outputs, a reasonable way is to study specific intraseasonal variability, and in this study, we focus on 30-80-day signals. However, due to the data length of forecast results, a band-pass filter method is not appropriate for extracting intraseasonal signals for FuXi-S2S and ECMWF S2S models. Instead, a non-filtering method proposed by Hsu et al. [37] is utilized here to extract observed and predicted intraseasonal variability in accordance with the following three steps. For observations: (1) remove the mean and the first three harmonics of the climatological annual cycle (2002-2016); (2) subtract the previous 40-day mean to eliminate signals with a period longer than 80 days; (3) apply a previous 15-day running mean to remove the high-frequency component (<30 days). For model outputs, the model climatology from 2002 to 2016 is first removed for FuXi-S2S and ECMWF S2S models respectively. And the next two steps are the same as in the observation process. Moreover, the low-frequency background (LFB) component (>80 days) is obtained by subtracting results after step 2 from the raw data.

## 3. Results and discussion

### *3.1. Improved MJO prediction skills of the FuXi-S2S model*

The MJO prediction skill is commonly evaluated using scalar metrics, such as the ACC and RMSE relative to the RMM index proposed by WH04. These metrics are calculated as functions of the forecast lead time [35,38]. Consistent with previous works, we apply the threshold ACC=0.5 for the skillful prediction [33,38,39]. As forecast lead time increases, prediction skills decline for both models. Remarkably, the FuXi-S2S model outperforms the ECMWF S2S model most of the time, extending the skillful prediction to 35 days-7 days longer than the ECMWF S2S model's 28 days (Fig. 1a). And we further assess the forecast skill based on initial strong cases (RMM amplitude ≥ 1) and weak cases (RMM amplitude < 1). Similar to results for all cases, the FuXi-S2S model again shows superior performance compared to the ECMWF S2S model. It is worth mentioning that both models have higher skills for strong cases than for weak cases during the initial forecast period, but this relationship reverses in later forecast days (Fig. 1b, c). Additionally, we calculate skillful durations for each of the eight MJO phases. On average, skillful durations are 34 and 28.4 days, respectively for FuXi-S2S and ECMWF S2S models, with the FuXi-S2S model extending skillful prediction across all phases (Fig. 1d). For strong cases, the FuXi-S2S model has advantages in most phases except for phases 3, 5 and 8 (Fig. 1e). Regarding weak cases, the FuXi-S2S model only shows inferior performance in the phase 7 (Fig. 1f).

Meanwhile, in terms of the RMSE, forecasting skills for two models both show upward trends. Although the FuXi-S2S model generally exhibits lower RMSE than the ECMWF S2S model, their performance is comparable at certain times (Fig. 2a). For strong and weak cases, relationships between two models are consistent with that for all cases (Fig. 2b, c). The average RMSE during week 4 is also examined. For all cases, the forecast skill is comparable for the phase 1. Moreover, the FuXi-S2S model performs worse than the ECMWF S2S model in phases 3 and 8 (Fig. 2d). For strong cases, both models exhibit the largest RMSE in the phase 2. The FuXi-S2S model outperforms the ECMWF S2S model in phases 2, 4, 6, and 7 (Fig. 2e). For weak cases, the FuXi-S2S model consistently provides better results than the ECMWF S2S model except the phase 7 (Fig. 2f).

We also analyze the amplitude of the RMM index. For all cases or initial strong cases, predicted composite amplitudes decrease as forecast lead time increases, with the FuXi-S2S model consistently performing better than the ECMWF S2S model (Fig. 3a, b). As for initial weak cases, the observed amplitude shows an upward feature as the time progresses, reaching its peak at day 40. The largest amplitudes for FuXi-S2S and ECMWF S2S models occur at day 16 and day 10 respectively (Fig. 3c). Moreover, averaged amplitudes for week 4 across eight individual phases are also evaluated, and biases between the FuXi-S2S model and observations are consistently smaller (Fig. 3d-f).

### 3.2. Physical explanation for enhanced predictions of the FuXi-S2S model

The previous analysis is based on statistical metrics. Another fundamental question is whether the FuXi-S2S model still exhibits better performance in the MJO propagation process for the specific intraseasonal variability, and if so, whether we can further provide a physical explanation. Here, we concentrate on the 30-80-day intraseasonal variability and adopt the strong RMM phase 3 for the composite. It is worth mentioning that evaluating diversities of classic IO MJO cases for S2S models is also important [39]. However, due to the limited duration of the evaluation period, the sample size is relatively small and we may explore this in future work.

Hovmöller diagrams of intraseasonal variabilities of the OLR and 850-hPa zonal wind, averaged over 5°S to 5°N, are presented (Fig. 4). In observations, OLR and zonal wind anomalies both represent canonical eastward propagation features from the tropical IO to the tropical WP (Fig. 4a), and two models could forecast eastward propagation features (Fig. 4b, c). However, during day 15 to day 20, the MJO convective center located over the tropical WP, predicted intensities of OLR anomalies over this region are mainly underestimated for both two models. Notably, biases of OLR anomalies between the FuXi-S2S model and observations are predominantly more minor than those between the ECMWF S2S model and observations (Fig. 4b, c). Consequently, an interesting question arises: Could the improved performance in predicting OLR anomalies over the tropical WP during this period for the FuXi-S2S model be attributed to capturing the fundamental physical process more accurately? It has been widely acknowledged that the MJO could be viewed as a moisture mode, and thus, we try to explain the results above by examining vertically integrated moisture anomalies from 1000 hPa to 300 hPa [40-42] (Fig. 5). Consistent with the OLR propagation feature, the observed moisture anomaly also propagates eastward (Fig. 5a). And we also examine vertical structures of the moisture anomaly and its tendency. For instance, at day 5, there is a salient rearward (westward and upward) tilt of the moisture below 500 hPa, and the center of positive tendencies leads the center of enhanced moisture anomalies in observations, consistent with the moisture mode theory [43,44] (Fig. 6a). And FuXi-S2S and ECMWF S2S models could both capture these structure features well (Fig. 6b, c). With the MJO propagation, during days 15-20 predicted moisture anomalies over the tropical WP region for FuXi-S2S and ECMWF S2S models are less sufficient than observations, yet moisture anomalies in the FuXi-S2S model are closer to observed values (Fig. 5b, c), which explain the better prediction of OLR anomalies over the tropical WP region during this period for the FuXi-S2S model.

Because we mainly focus on the area-averaged feature, temporal evolutions of area-averaged OLR and moisture anomalies on the intraseasonal timescale over the tropical WP region are presented (Fig. 7). For the OLR evolution, both models capture the transition feature from the positive phase to the negative phase in the first 20 days. Consistent with results in Fig. 4, negative OLR anomalies in observation during day 15 to day 20 are lower than forecasted signals for FuXi-S2S and ECMWF S2S models, accompanied by smaller biases for the FuXi-S2S model (Fig. 7a). And biases for moisture anomalies represent same results (Fig. 7b). Thus another interesting question arises: why does the FuXi-S2S model show a smaller moisture bias during day 15 to day 20? Since the moisture is closely linked to the moisture transport, we will try to explain this phenomenon from the perspective of moisture transport [44,45].

To address this question, the multi-scale interaction diagnosis associated with moisture transport is conducted to explore details of dynamic processes [6,43-45]. Specifically, we examine the interaction between the MJO-scale (30-80 days) and LFB (>80 days) signals. Since FuXi-S2S model outputs do not include the vertical motion, the advection of averaged LFB moisture by intraseasonal zonal and meridional wind anomalies, as well as the advection of the intraseasonal moisture anomaly by averaged LFB zonal and meridional winds are explored. To make statements concise, we adopt abbreviations, including ZIL, ZLI, MIL and MLI, to represent the four terms above, where the first, second and third letters refer to the direction, the time scale of the wind and specific humidity individually. It is worth mentioning that the averaged LFB moisture refers to the averaged moisture over the entire forecast period for all initial strong MJO phase 3. Temporal evolutions of these four dynamic terms linked to the moisture transport averaged over the tropical WP region are shown (Fig. 8a-d). For ZIL and ZLI terms, signals are larger for the ECMWF S2S model than the FuXi-S2S model in the first 20 days (Fig. 8a, b). Regarding the MIL term, results are significant in the first two weeks based on the observation. And in the first 10 days, observed positive signals are consistently larger than predicted results for both models, with the FuXi-S2S model's values being closer to observations (Fig. 8c). Additionally, the MLI term from day 5 to day 15 presents the same relationship as the MIL term we propose above (Fig. 8d). Subsequently, we focus on the moisture development stage (day 1 to day 10) and explore the relative contributions of four factors above to the improved performance in moisture prediction for the FuXi-S2S model. Results show that the more accurate prediction of moisture during day 15 to day 20 could be effectively attributed to the MIL, MLI and ZIL terms (Fig. 8f).

Given the largest improvement in the MIL term, we further investigate this specific term to address another crucial question: Why does the FuXi-S2S model show an improvement in the MIL term prediction? To explore these, spatial structures of meridional gradients of LFB moisture averaged over the entire forecast period, as well as intraseasonal meridional wind anomalies averaged from day 1 to day 10 from 1000 hPa to 300 hPa are shown (Fig. 9). For the meridional gradient of LFB moisture, spatial configurations for both models are remarkably similar with the observation (Fig. 9a-c). For our focused tropical WP region, area-averaged gradients are -0.378, -0.351 and -0.274 for the observation, FuXi-S2S and ECMWF S2S model respectively (unit omitted), which means with the same wind (Fig. 9d), the more accurate prediction of the area-average meridional gradient of LFB moisture over the tropical WP is beneficial for moisture prediction improvements during day 15 to day 20. Meanwhile, horizontal structures of intraseasonal wind anomalies for two models represent similar features with the observation, wind anomalies averaged over the tropical WP region are 0.224, 0.251 and 0.273 individually (unit omitted), contributing positively to the smaller MIL bias for the FuXi-S2S model we have identified above (Fig. 8f). In summary, the more accurate forecast of the meridional gradient of LFB moisture averaged over the tropical WP over the entire forecast period for the FuXi-S2S model is a pivotal factor for improving the prediction of moisture and OLR anomalies averaged over the tropical WP region during day 15 to day 20 (Fig. 7a, b).

### 3.3. Discussion

We compared prediction capabilities of the MJO during boreal winter for FuXi-S2S and ECMWF S2S models, finding that the FuXi-S2S model extends the MJO's skillful prediction window from 28 days to 35 days. For RMSE and RMM amplitude, the FuXi-S2S model also outperforms the ECMWF S2S model.

Although for these statistical indexes, the FuXi-S2S model's results are better, there is an urgent need to provide a physical explanation to help us understand model results well. Thus, we select the RMM phase 3 with the amplitude $\geq 1$ to composite. Interestingly, the FuXi-S2S model demonstrates reduced biases of intraseasonal OLR anomalies averaged over the tropical WP region during days 15-20, along with more accurate area-averaged intraseasonal vertically integrated moisture anomalies over this region. We have conducted the multi-scale interaction diagnosis associated with moisture transport to provide a fundamental physical explanation. Results underscore that the more accurate moisture prediction could be attributed to the MIL term. And the improvement in the MIL term could be further effectively by the meridional gradient of the LFB moisture. The above analysis provides us with a reasonable physical explanation from the moisture mode perspective, helping us understand the better performance of the FuXi-S2S model.

However, our analysis only focused on the prediction capability of the MJO during boreal winter. Further investigation is needed to assess whether the FuXi-S2S model offers superior performance for the boreal summer intraseasonal oscillation (BSISO) and whether a similar physical explanation can be provided. Moreover, as we all know, extratropical circulation anomalies on the intraseasonal timescale are determined by the MJO convection to some extent, it remains an open question whether the FuXi-S2S model maintains the improved performance in this regard. Given that the current evaluation period is limited to 5 years, exploring the diversity of MJO/BSISO events is not feasible at this stage, and we plan to extend the evaluation period in future studies to address these questions.

## Author contributions

CC and XZ designed this research under the guidance of supervisors ZW and HL. CC made the data analysis and graphics. CC and XZ wrote the first draft and all authors contribute to discussion and comments on the manuscript.

## Competing interests

Authors declare that they have no competing interests.

## Data and materials availability:

Interpolated daily OLR data and ERA reanalysis dataset were obtained from websites (https://psl.noaa.gov/data) and (https://cds.climate.copernicus.eu/) respectively. ECMWF reforecasts were obtained from https://apps.ecmwf.int/datasets/data/s2s/. As the FuXi-S2S model and data are essential resources for this study. Currently, access to these resources is limited. For inquiries please contact HL at the following email address: lihao_lh@fudan.edu.cn.

## Acknowledgements:

We really acknowledge the useful discussion with B. Wang on an early stage of our work. **Funding**: This research was jointly supported by National Natural Science Foundation of China (NSFC) Major Research Plan on West-Pacific Earth System Multi-spheric Interactions (grant no. 92158203), the Ministry of Science and Technology of China (grant no. 2023YFF0805100), the Second Tibetan Plateau Scientific Expedition and Research (STEP) program (grant no. 2019QZKK0102), and the University of Sydney-Fudan University Ignition Grants.

# Figures

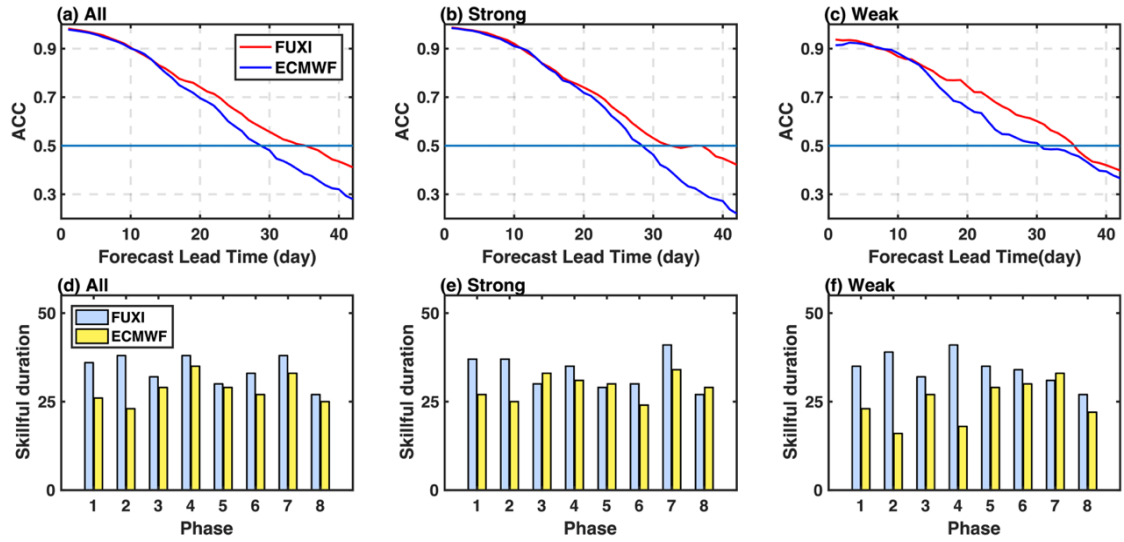

**Fig. 1.** Evolutions of the anomaly correlation coefficient about the Real-time Multivariate MJO (RMM) index between model results and observations with the forecast lead time during boreal winter. (a-c) total cases, initial strong cases (RMM amplitude ≥ 1, the same as below) and weak cases (RMM amplitude < 1, the same as below) respectively. Skillful duration (unit: day) corresponding to each of eight RMM phases for the FuXi-S2S model and the ECMWF S2S model. (d-f) total cases, initial strong cases and weak cases individually. Red and blue lines in (a-c), and blue and yellow bars in (d-f) indicate results of FuXi-S2S and ECMWF S2S models respectively.

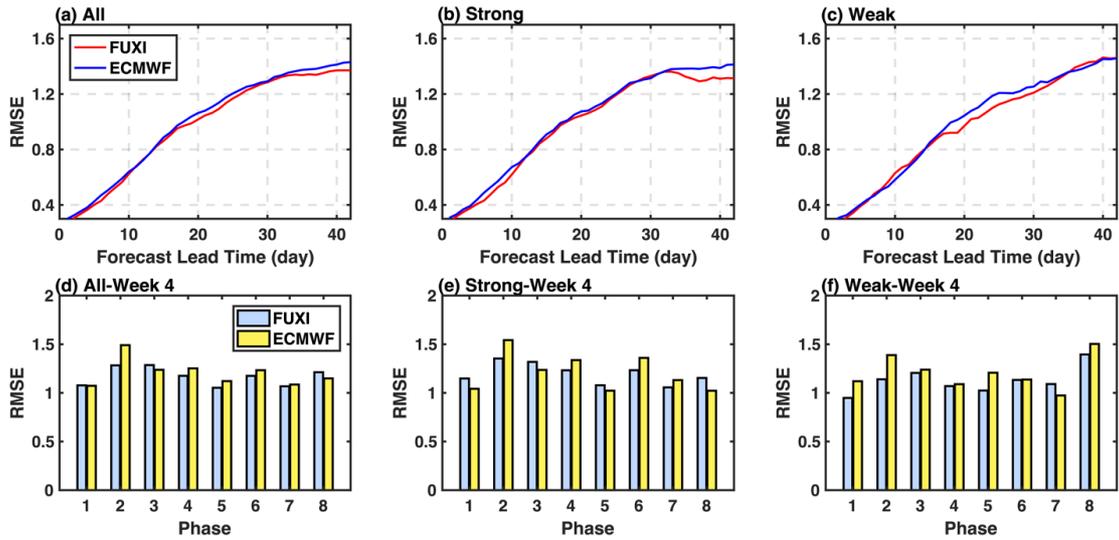

**Fig. 2.** Evolutions of the root-mean-square error (RMSE) for the RMM index between model results and observations with the forecast lead time during boreal winter. (a-c) total cases, initial strong cases and weak cases respectively. Averaged RMSE in the week 4 corresponding to eight individual phases for FuXi-S2S and ECMWF S2S models. (d-f) total cases, initial strong cases and weak cases individually. Red and blue lines in (a-c), and blue and yellow bars in (d-f) indicate results of FuXi-S2S and ECMWF S2S models respectively.

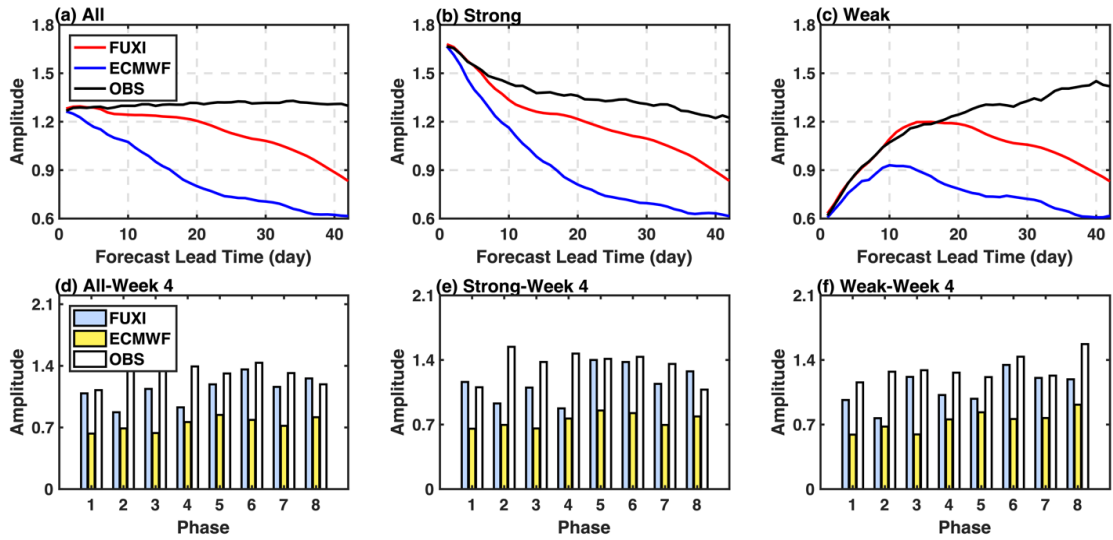

**Fig. 3.** Evolutions of the amplitude of the RMM index for the FuXi-S2S model, ECMWF S2S model and observations with the forecast lead time during boreal winter. (a-c) total cases, initial strong cases and weak cases respectively. Averaged amplitude in the week 4 corresponding to eight individual phases for FuXi-S2S model, ECMWF S2S models and observations. (d-f) total cases, initial strong cases and weak cases individually. Red, blue and black lines in (a-c), and blue, yellow and white bars in (d-f) indicate results of the FuXi-S2S model, the ECMWF S2S model and observations respectively.

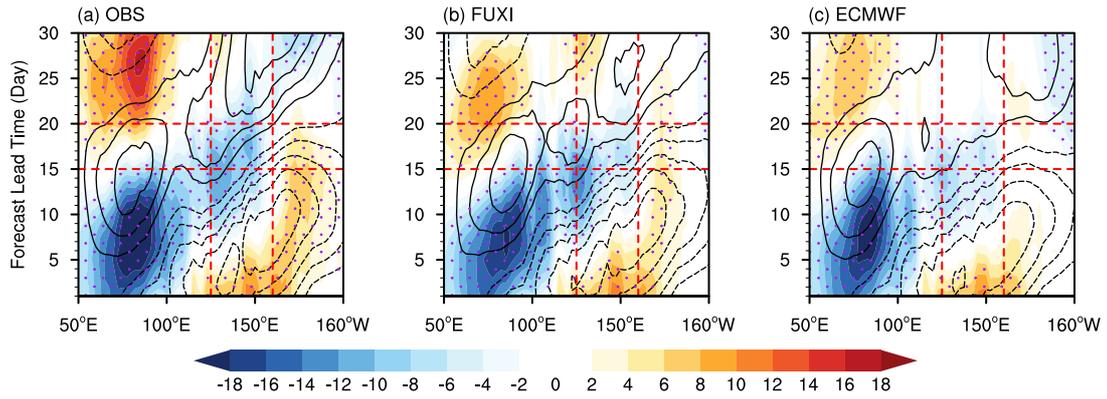

**Fig. 4.** Hovmöller diagrams of composite intraseasonal outgoing longwave radiation (OLR) anomalies (shading, unit: W m$^{-2}$) and 850-hPa zonal wind anomalies (contour with an interval of 0.5, unit: m s$^{-1}$) averaged over 5°S-5°N at the strong RMM phase 3 during boreal winter for the (a) observation, (b) FuXi-S2S model and (c) ECMWF S2S model. Solid and dashed contours in (a-c) indicate positive and negative values respectively. The zero line is omitted. Red lines indicate focused forecast period during day 15 to day 20 and tropical WP region (5°S-5°N, 125°-160°E). Dots indicate results pass the significant test at the 90% confidence level.

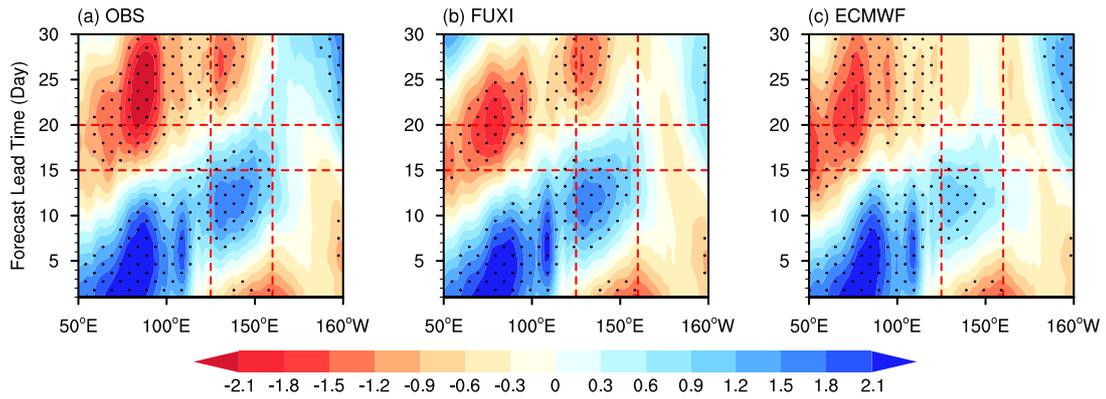

**Fig. 5.** Hovmöller diagrams of composite intraseasonal vertically integrated moisture anomalies from 1000 to 300 hPa (unit: kg m$^{-2}$) averaged over 5°S-5°N at the strong RMM phase 3 during boreal winter for the (a) observation, (b) FuXi-S2S model and (c) ECMWF S2S model. Red lines in (a-c) indicate focused forecast period from day 15 to day 20 and tropical WP region. Dots indicate results pass the significant test at the 90% confidence level.

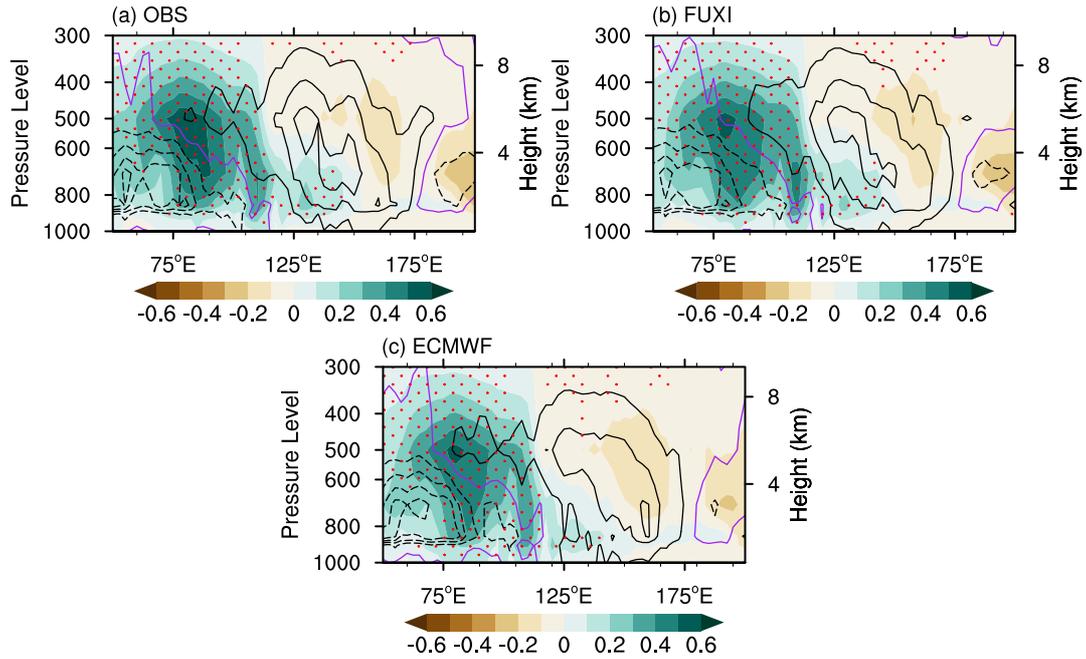

**Fig. 6.** Vertical structures of composite tropical (5°S-5°N) intraseasonal moisture anomalies (shading, unit: g kg$^{-1}$) and associated moisture anomaly tendencies (contour with an interval of 0.02, unit: g kg$^{-1}$ day$^{-1}$) at 5 days lag of the strong RMM phase 3 during boreal winter for the (a) observation, (b) FuXi-S2S model and (c) ECMWF S2S model. Dots indicate signals of moisture anomalies pass the significant test at the 90% confidence level. Solid and dashed contours indicate positive and negative values respectively. The purple line means the zero line.

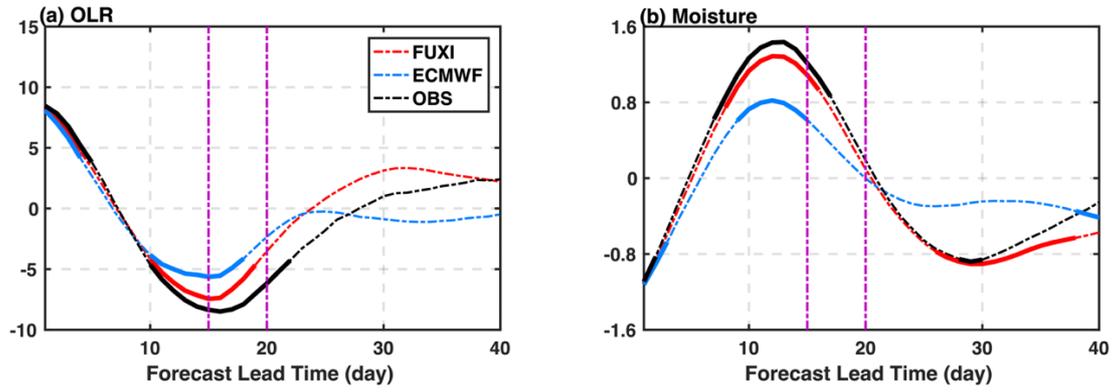

**Fig. 7.** Temporal evolutions of area-averaged intraseasonal OLR anomalies (a, unit: W m$^{-2}$) and moisture anomalies (b, unit: kg m$^{-2}$) over the tropical WP region for the strong RMM phase 3 during boreal winter. Red, blue and black lines in (a, b) indicate results of the FuXi-S2S model, ECMWF S2S model and observation respectively. Purple lines indicate day 15 and day 20.

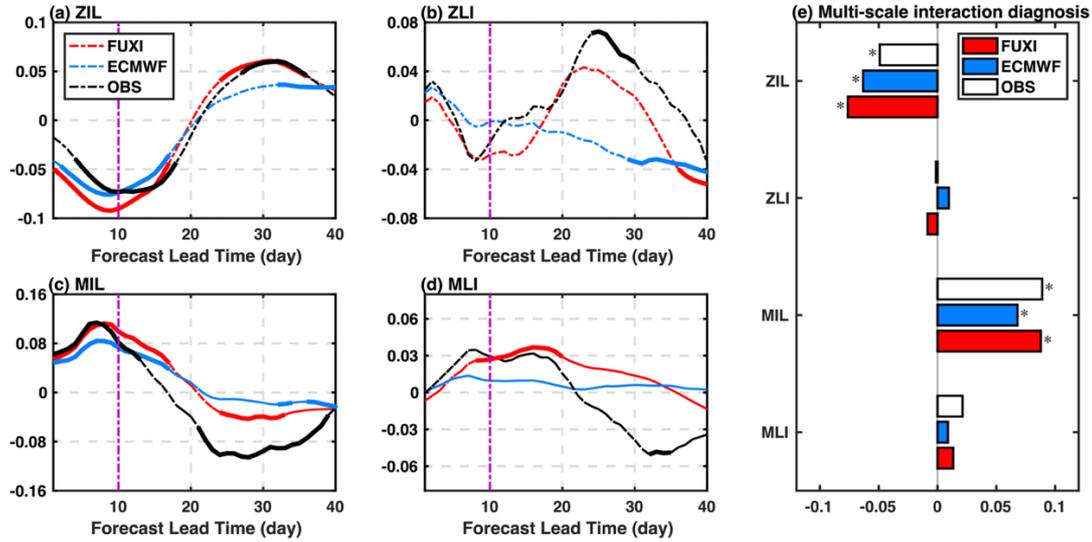

**Fig. 8.** Temporal evolutions of area-averaged multi-scale interaction terms over the tropical WP region. (a-d) advection of averaged low-frequency background (LFB) moisture by the intraseasonal zonal wind anomaly (unit: kg m$^{-2}$ day$^{-1}$), advection of the intraseasonal moisture anomaly by averaged LFB zonal wind (unit: kg m$^{-2}$ day$^{-1}$), advection of averaged LFB moisture by the intraseasonal meridional wind anomaly (unit: kg m$^{-2}$ day$^{-1}$) and advection of the intraseasonal moisture anomaly by averaged LFB meridional wind (unit: kg m$^{-2}$ day$^{-1}$) integrated from 1000 to 300 hPa respectively. Four dynamic terms averaged from day 1 to day 10 (e, unit: kg m$^{-2}$ day$^{-1}$) are also shown. Red, blue and black lines in (a-d), and red, blue and white bars in (e) indicate results of the FuXi-S2S model, ECMWF S2S model and observation individually. Thickened solid parts of lines in (a-d) and asterisks in (e) mean results pass the significant test at the 90% confidence level. Purple lines in (a-d) indicate day 10.

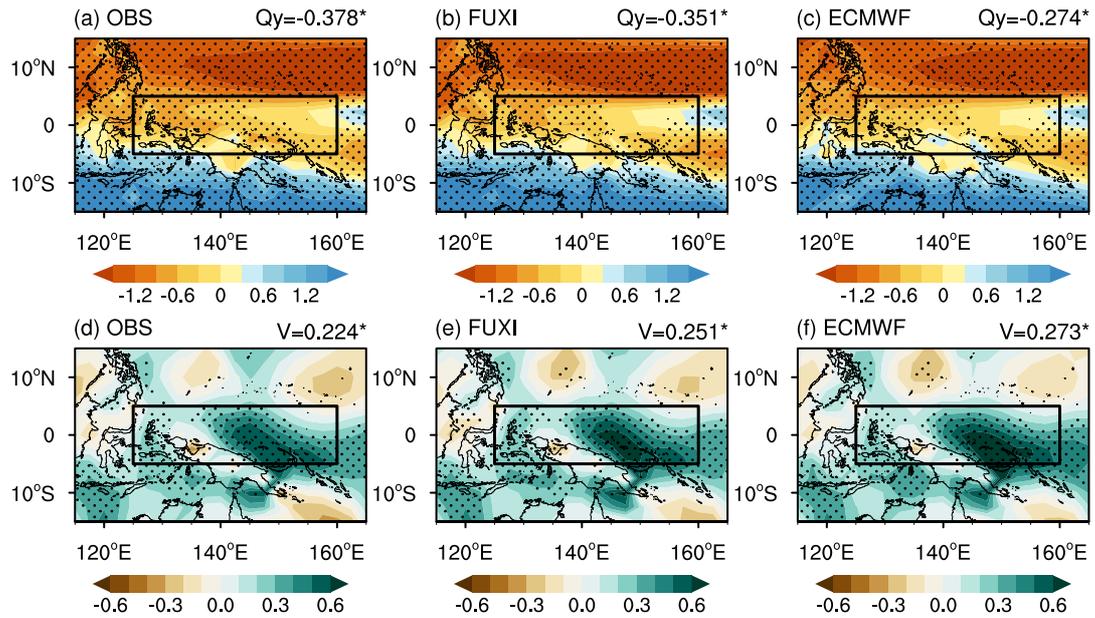

**Fig. 9.** Spatial structures of meridional gradients of LFB moisture (unit: $10^{-9}$ kg kg$^{-1}$ m$^{-1}$) averaged during the entire forecast period from 1000 to 300 hPa for the (a) observation, (b) FuXi-S2S model and (c) ECMWF S2S model. (d-f) As in (a-c), but for intraseasonal meridional wind anomalies (unit: m s$^{-1}$) averaged from day 1 to day 10 from 1000 to 300 hPa. Dots and asterisks indicate results pass the significant test at the 90% confidence level. Black boxes show the focused tropical WP region.